\documentclass{article}
\usepackage{spconf,amsmath,graphicx}
\usepackage{hyperref,float}

\title{A Language model Based Approach towards large scale and lightweight language identification systems}
\name{Brij Mohan Lal Srivastava$^{\dagger}$ \qquad Hari Krishna Vydana$^{\dagger}$ \qquad Anil Kumar Vuppala$^{\dagger}$ \qquad Manish Shrivastava$^{\star}$}

\address{International Institute of Information Technology, Hyderabad, India \\
	$^{\dagger}$ Speech \& Vision Lab\\ $^{\star}$ Language Technology Research Center\\
	\{brijmohanlal.s, hari.vydana\}@research.iiit.ac.in\\ \{anil.vuppala, m.shrivastava\}@iiit.ac.in}

\begin{document}
\maketitle
\begin{abstract}
Multilingual spoken dialogue systems have gained prominence in the recent past necessitating the requirement for a front-end Language Identification (LID) system. Most of the existing LID systems rely on modeling the language discriminative information from low-level acoustic features. Due to the variabilities of speech (speaker and emotional variabilities, etc.), large-scale LID systems developed using low-level acoustic features suffer from a degradation in the performance. In this approach, we have attempted to model the higher level language discriminative phonotactic information for developing an LID system. In this paper, the input speech signal is tokenized to phone sequences by using a language independent phone recognizer. The language discriminative phonotactic information in the obtained phone sequences are modeled using statistical and recurrent neural network based language modeling approaches. As this approach, relies on higher level phonotactical information it is more robust to variabilities of speech. Proposed approach is computationally light weight, highly scalable and it can be used in complement with the existing LID systems.       
\end{abstract}
\begin{keywords}
Language Identification, Recurrent neural network language model (RNNLM), SRI language model (SRILM), Phone recognizer followed by language model (PRLM), phonotactics.
\end{keywords}
\section{Introduction}
\label{sec:intro}

Language identification (LID) refers to the task of automatically identifying the language from the speech utterance. An LID
system is a vital module for a wide range of multilingual applications like, call centers, multilingual Spoken Dialog Systems, emergency services and speech-to-speech translation systems. Human-Computer interaction through speech can be taken more deeply into human society if the interaction is through multiple regional languages, for that LID system is a preliminary requirement. A lot of scientific interest is being shown in developing an LID system, specifically a lightweight system to lower the overhead cost. In relevance to the task of developing an LID system, recent works have focused on developing algorithms to extract suitable features for language identification. Lately i-vector based features are explored and they have exhibited better performance compared to the conventional spectral features like Mel-frequency cepstral coefficients (MFCC), Linear predictive cepstral coefficients (LPCC) and Shifted delta cepstral coefficients (SDC) in NIST evaluations for speaker and language recognition tasks~\cite{mccree2015dnn}. Neural networks have also been employed as feature extractors to compute the stacked bottleneck features for LID~\cite{matejka2014neural}. Multilingual bottleneck, multilingual tandem bottleneck obtained by stacking the SDC with the corresponding bottleneck features are explored in~\cite{fer2015multilingual,geng2015multilingual}. In most of the recent approaches the power of DNN (Deep neural networks) is explored for the task of language identification. Various approaches featuring Feed Forward Deep Neural Networks (FF-DNNs) and Long-Short Term Memory Recurrent Neural Networks (LSTM-RNNs) have been employed for developing the language classifiers. Additionally, convolutional neural networks have been studied to develop an end-to-end LID system for 8 languages in\cite{lozano2015end}.\\

The approaches mentioned above rely on the language discriminative capability present in the lower-level acoustic information with some contextual information in time neighborhood. The better performance of i-vector based approaches compared to the conventional spectral features can be attributed to its better context modeling capability. The systems which try to model the language discriminative information at higher-level (phones, phone frequency and phonotactics) have exhibited better performance~\cite{zissman2001automatic}. In this paper, we propose an approach to capture higher-level language discriminative information, which can be used as a complementary information to the existing low-level acoustic information modeling LID systems. A language independent phone recognizer followed by a language dependent phone model (PRLM) for capturing the phonotactics of 4 languages is used in~\cite{zissman1994ngram}. In~\cite{zissman1994ngram}, language dependent phone recognizer for every language operated in parallel (PPRLM) is used to decode the test utterance and the language model with large number of uni-gram and bi-gram counts is used as an indication to the spoken language identity. Though PPRLM (Parallel-Phone recognizer followed by language model) is quite efficient compared to PRLM, it is computationally inefficient to decode the test utterance using all language's phone recognizer. Scalability is the major issue for a PPRLM system whenever a new language has to be incorporated into the existing LID system.\\

In this paper, we attempt to explore the significance of PRLM based approaches for developing a large scale LID system. As a part of our experiments, we develop an LID system comprising of $176$ different languages. The pipeline of our system includes a language independent phone recognizer to generate the phone sequences from raw signal and two different language modeling approaches (SRILM and RNNLM) to capture the phonotactic information. The proposed approach is computationally quite efficient hence can be quite handy to add complementary evidence to the existing approaches. As the proposed approach mostly relies on the large durational phonotactic information of a language, it is more robust compared to low-level acoustic modeling approaches. The rest of the paper is organized as follows: Section 2 describes the dataset used in this approach. The proposed approach as such is described in section 3. Results and relevant discussions are presented in section 4. Conclusion and future scope are discussed in section 5.

\section{Data Description}
\label{sec:print}

The language data which is made openly available by Topcoder \footnote{\url{https://community.topcoder.com/tc?module=MatchDetails&rd=16555}} as a part of Spoken Language Recognition challenge is used in this work. Data sets comprises of recorded speech in 176 languages. The dataset contains $375$ utterances per language and the language labels of these utterances are also available. Each utterance has a duration of 10 seconds. Each speech recording is given in a separate file and only one language is spoken in each file. The available data is reorganized into training, testing, and validation sets. For training SRILM n-grams, $330$ utterances were used for developing language models and the remaining $45$ utterances are used for testing the models. In case of RNNLMs, $300$ utterances were used for training, $30$ utterances were used for validation and $45$ utterances are used to test the developed models. The data provided has speech recordings in mp3 format, which are later converted to WAV format with sampling rate as 16 kHz.\\

\section{Proposed Approach}
\label{sec:prior}
\begin{figure*}
\begin{center}
\includegraphics[scale=0.5]{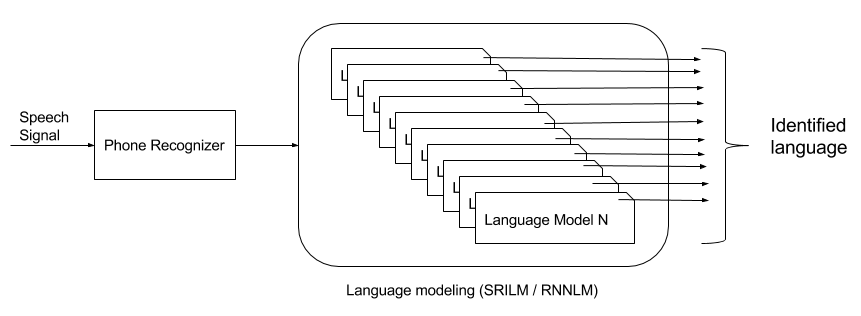}
\end{center}
 \caption{Block diagram of the proposed method.}
 \label{Block_diagram}
 \end{figure*}

In this approach, we mostly rely on the higher-level phonotactic information extracted from speech for developing an LID system. To extract the higher level phonotactic information input speech signal is to be tokenized. For tokenizing the input speech a language independent phone recognizer is used and the phone sequence is obtained. Although the phone recognizer is independent of the language that is being decoded, we assume that the similar sounding acoustic patterns will be decoded as approximately the same phone labels. By using the language independent phone recognizer, we are relying on the consistency of the phone recognizer rather than the accuracy i.e., similar sounding acoustic sounds will be tagged with same phone label regardless of the language. In this work, we hypothesize that the statistical patterns present in the obtained phone sequence have the language-discriminative information. For that purpose, SRILM and RNNLM are explored to model the statistical patterns that convey the language discriminative information from the tokenized phone sequence. The block diagram of the proposed approach is presented in Fig~\ref{Block_diagram}. The details of the phone recognizer and the language modeling techniques employed are described in the following subsections.\\

\subsection{Language-independent phone recognizer}
The goal of the language-independent phone recognizer is to provide maximum coverage of phone units present in all the languages for which the system is being developed. We employ PocketSphinx\cite{huggins2006pocketsphinx} as the front-end phone recognizer which uses HMM-based phone decoder from speech signal. A phonetically tied-mixture (PTM) model is used for efficient decoding. It contains 256 mixture components per state and assigns different mixture weights to the shared states of tri-phones. This model provides a good balance between speed and accuracy. Since it can be trained over huge data, it gives a decent decoding result in under real time. We use US English phone set with 40 phones and an unbiased phonetic language model for decoding.
The phone recognizer can be improved by training over multiple languages which will certainly increase the coverage of common phonetic and acoustic patterns. Current acoustic model is trained on US English speech data.

\subsection{Language modeling}
SRILM n-grams (uni-gram to 6-grams) and RNNLM have been used for experimentations to model the statistical patterns in the phone sequences.

\begin{figure}[H]
 \includegraphics[scale=0.5]{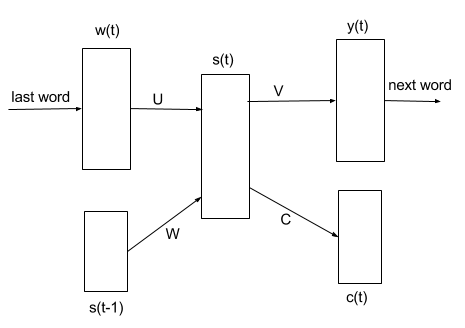}
 \caption{Architecture of RNNLM used in this work.}
 \label{rnnlm_arch}
\end{figure}

RNNLM as shown in Figure \ref{rnnlm_arch}, uses the current phone token $w(t)$ and previous state of hidden layer $s(t-1)$ to predict the probability of next token $y(t)$. The neuron in hidden layer $s(t)$ uses sigmoid activation function. Once the network is trained, we can use the output layer $y(t)$ as the probability distribution of the next word given current word and the state of hidden layer. Here, $c(t)$ represents the class layer which can be optionally used to reduce the computational complexity of model. As observed in the experiments conducted by us, models with lesser classes generally perform better at a higher computation cost. The matrix $W$ represents recurrent weights of the network which is trained using backpropagation through time (BPTT) algorithm. Training of $W$ can be improved by choosing optimum number of steps to propagate the error back in time. This can be task-dependent and for good LID models this value can be $4$ to $6$.

\section{Results \& Discussion}
\label{sec:majhead}

\begin{table}[H]
\centering
\begin{tabular}{|c|c|c|c|}
\hline
{\bf System}      &  Min&Max&Avg Accuracy\\ \hline
SRILM 1-gram      & 43.79&60&46.53             \\ \hline
SRILM 2-gram      & 77.77&83.61&81.77           \\ \hline
SRILM 3-gram      & 84.07&88.88&85.45 \\ \hline
SRILM 4-gram      & 81.11&86.66&83.94          \\ \hline
SRILM 5-gram      & 77.77&86.66&83.15           \\ \hline
SRILM 6-gram      & 77.77&86.66&82.98  \\ \hline \hline
RNNLM 1-class     & 83.13&{\bf 93.33}&84.42\\ \hline 
RNNLM 6-classes & {\bf 84.44}&90.58&{\bf 87.69}        \\   \hline
RNNLM 100-classes & 82.22&89.44&85.38        \\   \hline
\end{tabular}
\caption{Language identification accuracies for each language model (explicit approach)}
\label{results_table}
\end{table}

Column 1 of Table.~\ref{results_table} are the various language models developed during the study. Column 2 specifies the performance of language model with minimum accuracy among all the 176 language models. Similarly, column 3 specifies the performance of language model with maximum accuracy among all the 176 language models. Column 4 is the average percentage of correctly detected testcases in all the 176 languages. Columns 2, 3 are intended to show that the performance of LID system is consistant across the languages. In~Table.~\ref{results_table}, row 2-7 are the performances using SRILM and rows 8-10 are the performaces obtained using RNNLM.  

Language models are estimated for each of the 176 languages. While training, we experimented with different classes (1, 6, 100) and sizes of the hidden layers (30, 40) in RNNLM. Model with 6 classes and 40 units in hidden layer gives best accuracy averaged over all languages. The maximum accuracy obtained using RNNLM is $93.33\%$ for Dangaleat language. Although the performance of RNNLM models for LID can be boosted at the cost of higher computational complexity, we observed that tri-gram models with low complexity can be used to achieve comparable results.\\

RNNLM is known to exhibit high sequence learning capabilities \cite{mikolov2011rnnlm} so the language-discriminative patterns from the phone sequences is captured to a good extent which can observed from the results of section \ref{sec:majhead}. We also notice that n-grams accuracy increases drastically when we go from uni-gram to trigram models and then comes down gradually while we move to 6-gram models. For certain languages n-grams even surpass the accuracy obtained from a 100-class RNNLM. This observation can be utilized in order to linearly interpolate the scores obtained from both models to boost the accuracy to maximum. For testing, sentence-level perplexity obtained from each model is compared and the language which gives lowest perplexity value is chosen as the most probable label.\\

Based on the conducted experiments, we observed that RNNLM performance increases when we use optimum number of classes to decompose the vocabulary, i.e., when number of classes are approximately equal to $\sqrt{|V|}$, where $|V|$ is the vocabulary size. In case of RNNLM 6-classes, error has been propagated to $4$ steps back in time, which lets the model predict the next word probability with the knowledge of higher dimensional features captured by the history. According to Table~ \ref{results_table}, this model gives highest average language identification performance of $87.69\%$ at a much higher computational cost. Rest of the models use simple RNN to model the language.

\section{Conclusion}
\label{sec:foot}
In this work, we have studied the scalability of PRLM based approaches for language identification. The proposed approach employs a language independent phone recognizer for tokenizing the input speech to phone sequences. We have explored the language modeling approaches such as SRILM and RNNLM for modeling the statistical patterns in the phone sequence. As the proposed approach relies on the phonotactic information this can be used as a complementary information to the approaches that rely on language information from low-level acoustic features. The proposed approach is highly computationally efficient and it can come handy to enhance many other approaches of LID systems. From the results it can be observed that both SRILM and RNNLM based language models have shown equally good performance for developing a large scale LID systems.  \\

We have developed LID systems using sentence-level probabilities and n-best scores obtained from various language models. Currently, the language models are developed independent of each other. Some of the future tasks can be to develop linearly and non-linearly interpolated language models for LID, synthesizing LID-specific acoustic models and Language model pruning. Less phonetic coverage could be a reason for low recognition accuracy of some languages hence more generalized phone recognizer with large phonetic coverage could be developed specifically for LID systems. Based on the statistics obtained from the language-specific language models, further pruning can be done.\\



\bibliographystyle{IEEEbib}
\bibliography{strings,refs1}

\end{document}